\documentclass[twocolumn, trackchanges]{aastex63}
\usepackage{lineno}

\usepackage{graphicx, amsmath}
\usepackage[caption=false]{subfig}

\usepackage{hyperref}
\usepackage{cleveref}
\usepackage{xspace}
\usepackage{rotating}
\usepackage{comment}

\newcommand{\lya}{Ly$\alpha$\ }

\newcommand{\lyaa}{Ly$\alpha$}

\newcommand{\OII}{[\ion{O}{2}]\xspace}

\newcommand{\hst}{\mbox{$HST$}}

\newcommand{\kmss}{km $\mathrm{s}^{-1}$}
\newcommand{\kms}{km $\mathrm{s}^{-1}$\xspace}

\newcommand{\escma}{ergs~s$^{-1}$~cm$^{-2}$~\AA$^{-1}$}

\graphicspath{{./}{figures/}}

\begin{document}

\title{Absorption Troughs of Lyman Alpha
Emitters in HETDEX}

\author[0000-0002-4974-1243]{Laurel H. Weiss}
\affiliation{Department of Astronomy, The University of Texas at Austin, 2515 Speedway Boulevard, Stop C1400, Austin, TX 78712, USA}

\author[0000-0002-8925-9769]{Dustin Davis}
\affiliation{Department of Astronomy, The University of Texas at Austin, 2515 Speedway Boulevard, Stop C1400, Austin, TX 78712, USA}

\author[0000-0002-8433-8185]{Karl Gebhardt}
\affiliation{Department of Astronomy, The University of Texas at Austin, 2515 Speedway Boulevard, Stop C1400, Austin, TX 78712, USA}

\author[0000-0002-5659-4974]{Simon Gazagnes}
\affiliation{Department of Astronomy, The University of Texas at Austin, 2515 Speedway Boulevard, Stop C1400, Austin, TX 78712, USA}

\author[0009-0003-1893-9526]{Mahan Mirza Khanlari}
\affiliation{Department of Astronomy, The University of Texas at Austin, 2515 Speedway Boulevard, Stop C1400, Austin, TX 78712, USA}

\author[0000-0002-2307-0146]{Erin Mentuch Cooper}
\affiliation{Department of Astronomy, The University of Texas at Austin, 2515 Speedway Boulevard, Stop C1400, Austin, TX 78712, USA}

\author[0000-0002-0302-2577]{John Chisholm}
\affiliation{Department of Astronomy, The University of Texas at Austin, 2515 Speedway Boulevard, Stop C1400, Austin, TX 78712, USA}

\author[0000-0002-4153-053X]{Danielle Berg}
\affiliation{Department of Astronomy, The University of Texas at Austin, 2515 Speedway Boulevard, Stop C1400, Austin, TX 78712, USA}

\author[0000-0003-4381-5245]{William P. Bowman}
\affiliation{ Department of Astronomy, Yale University, New Haven, CT 06511, USA}

\author[0000-0002-0885-8090]{Chris Byrohl}
\affiliation{Universit\"ats
Heidelberg, Zentrum f\"ur Astronomie, Institut f\"ur Theoretische
Astrophysik, Albert-Ueberle-Str. 2, 69120 Heidelberg, Germany}

\author[0000-0002-1328-0211]{Robin Ciardullo} \affiliation{Department of Astronomy \& Astrophysics, The Pennsylvania State University, University Park, PA 16802, USA} 
\affiliation{Institute for Gravitation and the Cosmos, The Pennsylvania State University, University Park, PA 16802, USA}

\author[0000-0002-7025-6058]{Maximilian Fabricius}
\affiliation{Max-Planck-Institut f\"ur Extraterrestrische Physik,
Gie{\ss}enbachstra{\ss}e 1, 85748 Garching, Germany}
\affiliation{Universit\"ats-Sternwarte M\"unchen, Scheinerstra{\ss}e 1, D-81679
München, Germany}

\author[0000-0003-2575-0652]{Daniel Farrow}
\affiliation{Centre of Excellence for Data Science,
Artificial Intelligence \& Modelling (DAIM),
University of Hull, Cottingham Road, Hull, HU6 7RX, UK}
\affiliation{E. A. Milne Centre for Astrophysics
University of Hull, Cottingham Road, Hull, HU6 7RX, UK}

\author[ 0000-0001-6842-2371]{Caryl Gronwall}
\affiliation{Department of Astronomy \& Astrophysics, The Pennsylvania State University, University Park, PA 16802, USA} 
\affiliation{Institute for Gravitation and the Cosmos, The Pennsylvania State University, University Park, PA 16802, USA}

\author[0000-0001-6717-7685]{Gary J. Hill} 
\affiliation{McDonald Observatory, The University of Texas at Austin, 2515 Speedway Boulevard, Stop C1402, Austin, TX 78712, USA}
\affiliation{Department of Astronomy, The University of Texas at Austin, 2515 Speedway Boulevard, Stop C1400, Austin, TX 78712, USA} 

\author[0000-0002-1496-6514]{Lindsay R. House}
\altaffiliation{NSF Graduate Research Fellow}
\affiliation{Department of Astronomy, The University of Texas at Austin, 2515 Speedway Boulevard, Stop C1400, Austin, TX 78712, USA} 

\author[0000-0002-8434-979X]{Donghui Jeong}
\affiliation{Department of Astronomy \& Astrophysics, The Pennsylvania State University, University Park, PA 16802, USA}
\affiliation{Institute for Gravitation and the Cosmos, The Pennsylvania State University, University Park, PA 16802, USA}

\author[0000-0001-5610-4405]{Hasti Khoraminezhad}
\affiliation{Institute for Multi-messenger Astrophysics and Cosmology, Department of Physics, Missouri University of Science and Technology, 1315 N Pine Street, Rolla, MO 65409, USA}

\author[0000-0002-0417-1494]{Wolfram Kollatschny}
\affiliation{Institut f\"ur Astrophysik and Geophysik, Universit\"at G\"ottingen, Friedrich-Hund Platz 1, D-37077 G\"ottingen, Germany}

\author[0000-0002-0136-2404]{Eiichiro Komatsu}
\affiliation{Max-Planck-Institut f\"{u}r Astrophysik, Karl-Schwarzschild-Str. 1, 85741 Garching, Germany}
\affiliation{Ludwig-Maximilians-Universit\"at M\"unchen, Schellingstr. 4, 80799 M\"unchen, Germany}
\affiliation{Kavli Institute for the Physics and Mathematics of the Universe (Kavli IPMU, WPI), University of Tokyo, Chiba 277-8582, Japan}

\author[0000-0002-6907-8370]{Maja Lujan Niemeyer}
\affiliation{Max-Planck-Institut f\"{u}r Astrophysik, Karl-Schwarzschild-Str. 1, 85741 Garching, Germany}

\author[0000-0002-6186-5476]{Shun Saito}
\affiliation{Institute for Multi-messenger Astrophysics and Cosmology, Department of Physics, Missouri University of Science and Technology, 1315 N Pine Street, Rolla, MO 65409, USA}
\affiliation{Kavli Institute for the Physics and Mathematics of the Universe (Kavli IPMU, WPI), University of Tokyo, Chiba 277-8582, Japan}

\author[0000-0001-7240-7449]{Donald P. Schneider}
\affiliation{Department of Astronomy \& Astrophysics, The Pennsylvania
State University, University Park, PA 16802, USA}
\affiliation{Institute for Gravitation and the Cosmos, The Pennsylvania State University, University Park, PA 16802, USA}

\author[0000-0003-2307-0629]{Gregory R. Zeimann}
\affiliation{Hobby Eberly Telescope, University of Texas, Austin, Austin, TX, 78712, USA}

\begin{abstract}
The Hobby-Eberly Telescope Dark Energy Experiment (HETDEX) is designed to detect and measure the redshifts of more than one million \lya emitting galaxies (LAEs) between $1.88 < z < 3.52$. In addition to its cosmological measurements, these data enable studies of \lya spectral profiles and the underlying radiative transfer. Using the roughly half a million LAEs in the HETDEX Data Release 3, we stack various subsets to obtain the typical \lya profile for the $z \sim 2-3$ epoch and to understand their physical properties. We find clear absorption wings around \lya emission, which extend $\sim 2000$ \kms both redward and blueward of the central line.  Using far-UV spectra of nearby ($0.002 < z < 0.182$) LAEs in the CLASSY treasury and optical/near-IR spectra of $2.8 < z < 6.7$ LAEs in the MUSE-Wide survey, we observe absorption profiles in both redshift regimes. Dividing the sample by volume density shows that the troughs increase in higher density regions. This trend suggests that the depth of the absorption is dependent on the local density of objects near the LAE, a geometry that is similar to damped Lyman-$\alpha$ systems. Simple simulations of \lya radiative transfer can produce similar troughs due to absorption of light from background sources by \ion{H}{1} gas surrounding the LAEs.
\end{abstract}


\section{Introduction}\label{sec:intro}
The Lyman alpha (\lyaa) emission line has long been regarded as a powerful tool with which to probe galaxies in the act of formation \citep{Partridge_1967}. As expected, \lya emitters (LAEs) have become one of the most heavily observed components of the high-redshift universe, being identified as early as $z > 7$ with the \textit{James Webb Space Telescope} \citep[e.g.,][]{Tang_2023}. The \lya emission line is useful because it is easily detectable via narrow band imaging or spectroscopy \citep[see][for a review]{Ouchi_2020}. As a result, ground-based observations can use LAEs to trace the large-scale structure of galaxies  from $z \sim 1.8$ to $z \sim 6.5$. Detecting this feature is the primary goal of the Hobby Eberly Telescope Dark Energy experiment (HETDEX, \citealt{Gebhardt_2021, Hill_2021}). Using LAEs, HETDEX aims to measure the $z \sim 2.4$ Hubble expansion rate, $H(z)$, and the angular diameter distance, $D_A(z)$, by identifying and 3D mapping over one million LAEs \citep{Gebhardt_2021}. 

Given the importance of LAEs in the study of the early universe, it is essential that we understand their physical properties and the physics behind their \lya emission. Since \lya is a resonance line, its radiative transfer is complicated and results in a variety of observed \lya spectral profiles \citep{Verhamme_2006}. The fate of \lya photons depends on the geometry and physical conditions of not only the interstellar medium (ISM), but also the circumgalactic medium (CGM) and the intergalactic medium (IGM). After its production via photoionization, a Ly$\alpha$ photon’s escape from the ISM is dependent on the amount and distribution of dust \citep[e.g.,][]{ Schaerer_2011, Finkelstein_2008, Scarlata_2009}, ISM kinematics \citep[e.g.,][]{Kunth_1998}, and the galaxy's neutral gas content and geometry \citep[e.g.,][]{Neufeld_1991, Hansen_2006, Jaskot_2014}. Beyond the ISM and out to the galaxy's virial radius, \lya photons resonantly scatter throughout the CGM. 

Observationally, the CGM has been detected in extended \lya emission as \lya halos \citep[e.g][]{Steidel_2010, Matsuda_2012, Wisotzki_2016}, as well as in \lya absorption from foreground/background galaxy pairs \citep{Steidel_2010}. Studies aimed at modeling \lya in the CGM suggest that the gas is multiphase with complex kinematics \citep[e.g.][]{Dijkstra_2012, Shen_2013, Chung_2019}. For example, inflows and outflows can have a strong effect on the \lya profile shape \citep[e.g.][]{Park_2021}. 

Outside the virial radius, photons must traverse the IGM. This reservoir of gas resides in filaments between galaxies, tracing structures on megaparsec (Mpc) scales. In the rest frame of the intervening gas, photons at the \lya wavelength are readily absorbed/scattered, and this has long been observed along lines of sight to bright quasars as the Lyman alpha forest \citep[e.g.][]{Gunn_Peterson_1965, Lynds_1971, Hu_1995, Slosar_2011}. The physical properties of the ISM, CGM, and IGM are naturally encoded within an observed \lya profile, although the constraints on their influence are not yet fully developed.

The HETDEX is an untargeted spectroscopic survey, using integral field unit (IFU) spectroscopy to detect and measure the redshift of \lya from $z\sim 1.88-3.52$ galaxies.  The more than 1 million LAE spectra expected from HETDEX provide the opportunity to investigate \lya emission in the context of a variety of properties. There have been a wealth of observations of \lya in both the local \citep[e.g.][]{Kunth_1998, Cowie_2010, Rivera-Thorsen_2015, Berg_2022} and high redshift universe \citep[e.g.][]{Ouchi_2008, Stark_2010, Shibuya_2012, Urrutia_2019}. Comparisons of HETDEX LAEs to spectra of LAEs at lower and higher redshifts provide an opportunity to investigate the profile of \lya in LAEs across cosmic time, and, by extension, to investigate key aspects of galaxy evolution. Furthermore, an understanding of the profiles of \lya helps HETDEX discriminate between true LAEs and false detections and thereby reduce the measurement error on $H(z)$ and $D_A(z)$. 

While the individual exposure time in HETDEX of $\sim 18$ minutes on source does not provide sufficient signal-to-noise ratio ($S/N$) to investigate the detailed structure of the \lya profile for an individual LAE, we can stack the spectra of the survey's $\sim 1$~million LAEs to greatly enhance the signal. Given the size and breadth of the sample, we can explore the stacks according to physical properties such as redshift, mass, and environment.

This paper is organized as follows. Section \ref{sec:data} introduces LAE datasets from HETDEX, the COS Legacy Archive Spectroscopic Survey (CLASSY; \citealt{Berg_2022}), and the Multi Unit Spectroscopic (MUSE)-wide survey \citep{Bacon_2010}. Section \ref{sec:stacking} presents the stacking methodology. Section \ref{sec:troughs} discusses the analysis of these stacks, and \ref{sec:origins} discusses a physical interpretation of our results. Throughout the paper, we will assume the Planck 2018 cosmology \citep{Planck_2018} with $\Omega_{\text{m}}$= 0.31 and $H_0$ = 67.7 $\mathrm{km~s^{-1}~Mpc^{-1}}$. All magnitudes are in the AB system \citep{Oke_Gunn_1983}.

\begin{figure*}
    \centering
    
    \includegraphics[height=5.8cm]{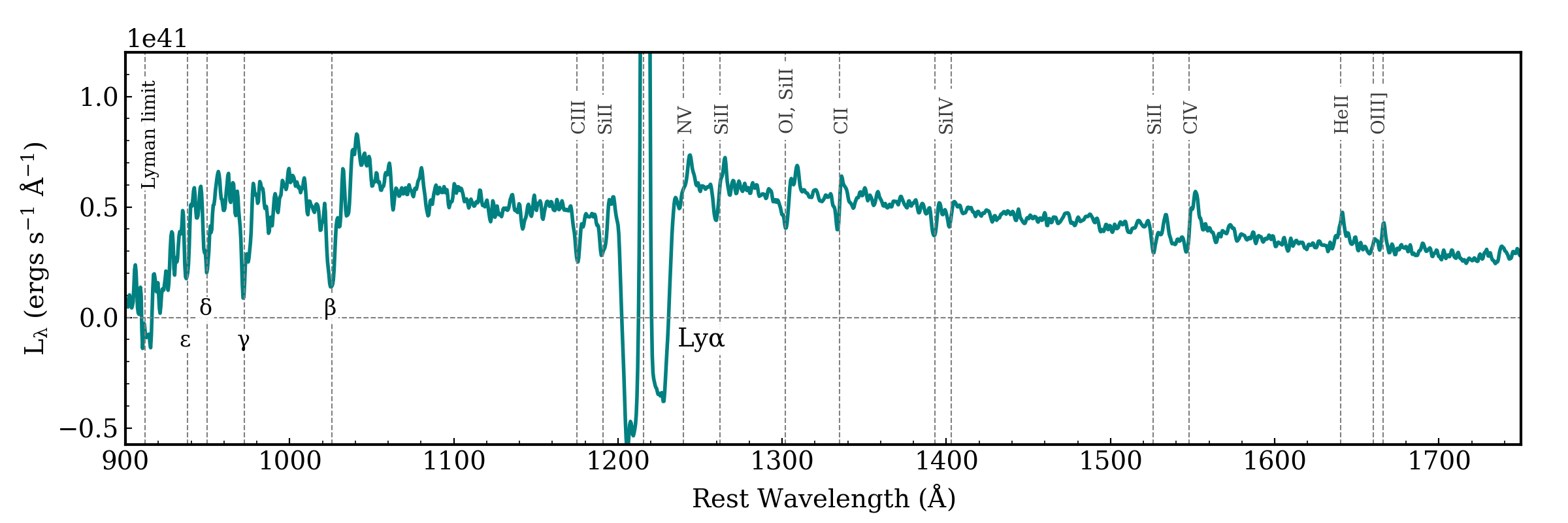}
    \caption{A stack of $\sim 50,000$ high-confidence LAE spectra from HPSC-1, as presented in \cite{Davis_2023_hpsc}. We select the spectra that have \lya linewidth $\sigma < 5.5$ \AA. Faint emission and absorption lines are marked and labeled. While \lya emission is detected in individual HETDEX spectra, the other faint emission/absorption features only become visible in stacks of $\gtrsim 10,000$ sources. The significantly negative flux values of the \lya absorption troughs are likely the result of background over subtraction as discussed in text, i.e. the absorption troughs are real, while their negative fluxes result from our assumptions in sky subtraction.}
    \label{fig:hpsc stack}
\end{figure*}

\section{Optical and UV Spectroscopy}
\label{sec:data}

The HETDEX \citep{Gebhardt_2021,Hill_2021} is a multi-year, untargeted spectroscopic survey conducted with the upgraded Hobby-Eberly Telescope \citep[HET,][]{Ramsey_1998, Hill_2021}.  The HETDEX instrument is the Visible Integral-Field Replicable Unit Spectrograph (VIRUS, \citealt{hil18a, Hill_2021}). VIRUS consists of 78 IFUs and 156 spectrographs, where each IFU covers $51\arcsec \times 51\arcsec$ on the sky. Each IFU contains 448 $1\farcs 5$-diameter optical fibers, which feed a pair of low-resolution ($750 < R < 950$) spectrographs covering the wavelength range between 3500~\AA\ and 5500~\AA. 
The fibers in a given IFU are spaced to give a 1$/$3 fill-factor,
so that a three position dithered set of exposures provides full spatial coverage within an IFU \citep{Hill_2021}. The typical exposure time of $\sim 18$ minutes over three dithers then provides 3$\times$34,944 spectra. 

The processing of HETDEX frames is described in detail in \cite{Gebhardt_2021}. The spectra are calibrated, sky subtracted, and inspected for emission lines and continuum sources. If an object is detected in a fiber, a point-spread-function (PSF) weighted spectrum is extracted from the surrounding fibers. The PSF-weighted spectra are then classified and their redshifts are determined with the ELiXer software package \citep{Davis_2023}. 

With the exception of the stack from the HETDEX Public Source Catalog (HPSC-1, see \citealt{Cooper_2023} and \citealt{Davis_2023_hpsc}), the LAE spectra in this project come from HETDEX Internal Data Release 3.0.1 (HDR3). This release contains all HETDEX data from January 3, 2017, up to and including August 3, 2021. The updated source catalog contains (at $S/N >5$) $\sim 520,000$ LAEs. In this paper, we select a set of $\sim 300,000$ spectra from this sample, which have high-confidence LAE classifications from ELiXer ($P_{\mathrm{Ly}\alpha} > 0.8$).  We then further restrict the LAE sample to objects with emission-line detections with $S/N > 5$ and linewidths of $\sigma < 5.5$~\AA\ ($\sim350$ \kms); this removes potential AGN that were not identified as such by EliXer or cataloged in \cite{Liu_2022}. 

To compare our LAE spectra to other surveys at low and high redshift, we make use of the CLASSY and MUSE-Wide surveys. \cite{Berg_2022} present the high-resolution ($R \sim 15,000$) far-UV (900-2000~\AA) CLASSY spectra and the high-level data products from the \textit{Cosmic Origins Spectrograph} (COS) on the \textit{Hubble Space Telescope} (\hst). The sample of 45 galaxies was chosen to be star forming, compact (near-UV FWHM $<$ 2\farcs5), low redshift ($z < 0.2$), and UV bright ($\rm{m_{FUV}} < 20$). For further discussion on data reduction, extraction, flux calibration, and co-addition of the CLASSY Treasury, see \cite{Berg_2022}. 

The MUSE \citep{Bacon_2010} integral-field unit spectrograph on the \textit{Very Large Telescope} produces $R \sim 2000 - 4000$ optical spectra (4700-9300 \AA) over a $1\arcmin \times 1\arcmin$ field-of-view. MUSE-Wide is a blind spectroscopic survey that covers $100 \times~1~\rm{arcmin}^2$ target fields, with a total survey volume in \lya that amounts to $\sim 10^6~\rm{Mpc}^3$. The first 44 pointings taken in the CANDELS/GOODS-S and CANDELS/COSMOS regions are presented in \cite{Urrutia_2019}. This sample consists of 479 non-AGN LAEs with \lya based redshifts $2.9 < z < 6.3$. For LAEs with cataloged photometric counterparts, they reported spectral energy distribution (SED)-determined mass estimates of $7.5 \lesssim$ log$(M / M_{\odot}) \lesssim 10$. More information about the MUSE-Wide observations and sample can be found in \cite{Urrutia_2019}.

\section{Spectral Stacking}
\label{sec:stacking}

Our goal is to use high $S/N$ spectra around the \lya line to explore the physical properties of LAEs. This can only be done by stacking; at $z \gtrsim 2$, individual Ly$\alpha$ detections lack the S/N required for such an analysis. Our input sample includes 300,000 LAEs over a redshift range $1.88 < z < 3.52$. The individual spectra are flux calibrated to about 5\% accuracy by the HETDEX project \citep{Gebhardt_2021}. However, at the extremely faint flux levels that stacking provides, we require additional calibrations to the baseline reductions, including both sky subtraction and flux corrections. The sky subtraction, additional flux calibrations, wavelength shifts, and stacking procedures are described below \citep[see also][]{Gebhardt_2021}. 

\subsection{Additional Corrections for Stacking Analyses}
\label{sec:corrections}

Prior to stacking, we perform several corrections to the processed spectra. As a first step, we use sky residuals to provide a better estimate of the sky. The sky subtraction procedure removes extraneous light from an object's spectrum that is due to the atmosphere, foreground/background sources, and/or instrumental effects. For a given observation, HETDEX measures the sky both using the full 21\arcmin\ diameter focal plane (all 34,944 fibers), and using the local fibers within the immediate vicinity of the detection. We use the latter for this work. This local sky subtraction is based upon an individual amplifier (112 fibers), corresponding to an on-sky region of $50\arcsec\times 12\farcs5$. We remove fibers with continuum that have more than $3\times$ the biweight scale \citep{Beers_1990} of the fibers on the amplifier. Significantly extended, diffuse flux associated with a source detection can be removed using the local sky subtraction. For a more detailed description of HETDEX sky subtraction, see \citep{Gebhardt_2021}. This procedure provides our initial sky-subtracted spectra for all fibers. 

Since the act of stacking pushes the flux limit of our data far below the level of the background sky, we must check sky residuals and apply any corrections to the initial analysis as described in \cite{Davis_2023_hpsc}. For each field, we generate 200 random aperture spectra from regions of the empty sky and stack them using a biweight \citep{Beers_1990}. We define our empty sky apertures such that 1) none of the 200 sky regions is within 1\farcs5 of each other, 2) each region contains at least 15 fibers, 3) the measured $g$ magnitude in the PSF-weighted spectra is fainter than 24, and 4) there are no detected sources within 2\arcsec. This residual stack is subtracted from the observed frame fiber spectra of a detection. This subtraction accounts for small calibration corrections, including flux from faint background and foreground sources. For an individual spectrum, this correction is less than 1\% of the original sky, far below the typical noise level, and the effect only becomes significant when stacking hundreds or thousands of HETDEX spectra.

After applying this correction, the observed air wavelengths are corrected to vacuum via \cite{Greisen_2006}, and the spectra are shifted to the rest frame using their redshifts as determined by HETDEX \citep{Cooper_2023}. Since the reported redshifts are measured from \lya alone, they do not account for any \lya velocity offset from the source's systemic redshift. For parts of this paper, we only correct the wavelengths for redshift, but do not apply any further corrections to the fluxes (such as cosmological dimming, Milky Way dust extinction, etc.). We will discuss the reasons for this approach shortly. We do note that stacking observed fluxes at different redshifts slightly biases the stack towards lower redshift objects, as comparatively more flux from these sources contributes, due to less cosmological dimming. Thus, when possible, we convert the observed flux densities to luminosity densities prior to stacking to reduce this bias.  

\subsection{Stacking Methodology}
By stacking hundreds to thousands of spectra, we significantly increase the $S/N$ of detections, and greatly improve our ability to identify and measure weak spectral features. In the case of HETDEX spectra, the stacking of LAEs creates  an ``average'' \lya profile with an increase in $S/N$ that is roughly proportional to the square root of the number of sources. Moreover, by creating sub-samples of galaxies based on the system's physical properties, we can investigate the systematics of \lya as a function of parameters such as continuum magnitude and emission-line strength. Since HETDEX is an untargeted survey with hundreds of thousands of spectra currently available, sufficiently large stacks of LAEs average over orientations, environment, geometries, and lines of sight.

We use the stacking method described in \cite{Davis_2021} and \cite{Davis_2023_hpsc}. The extent of the restframe wavelength coverage is determined from the highest and lowest redshift objects to be stacked, with a grid spacing adopted from the highest redshift object ($0.44$~\AA~for $z=3.5$). We linearly interpolate the individual restframe spectra onto this grid and stack each wavelength bin using a weighted biweight. The weighting scheme is a modified version of Tukey's biweight estimator \citep{Andrews_1972, Beers_1990} where each point is weighted by the inverse of the flux variance \citep{Davis_2021}. For large stacks of more than $\sim 1000$ spectra, using the mean, median, biweight, and weighted biweight statistics show little difference \citep{Davis_2023_hpsc}.

Figure \ref{fig:hpsc stack} depicts the stack of $\sim 50,000$ HETDEX LAEs from HPSC-1 presented in \cite{Davis_2023_hpsc}. One of the most prominent features in the spectrum are the significant absorption troughs on either side of the \lya emission. These troughs, as well as their properties and possible physical origin, are discussed in the following sections. Aside from \lyaa, many other spectral features are noticeable in the average stack. While \lya is visible in individual HETDEX spectra, faint emission and absorption lines, including the Lyman series, become apparent in a stack of many spectra. For a further detailed analysis of this stack, including the SED fitting and derived properties, see \cite{Davis_2023_hpsc}. 

\section{\texorpdfstring{\lyaa}{Lyman Alpha} Absorption Troughs}
\label{sec:troughs}
Perhaps the most striking feature in Figure \ref{fig:hpsc stack} is the deep absorption troughs blue-ward and red-ward of the \lya emission. The existence of both \lya absorption and emission in the spectra of $z \sim 2.5$~LAEs is not unexpected: for example, \cite{Erb_2014, Trainor_2015} and \cite{Trainor_2019} have all observed LAEs with \lya absorption primarily blueward of the emission (see Figures 3, 19, and 1 of these studies, respectively). Another example of this combined \lya profile is the damped \lya system (DLA) at $z = 2.0395$ presented in \cite{Moller_2004}. As in the case of the stacked spectrum shown in Figure \ref{fig:hpsc stack}, the 2D spectrum of the DLA displays \lya emission directly centered in an absorption well (see Figure 1 of \citealt{Moller_2004}). We will revisit DLAs in the context of HETDEX LAEs in Section \ref{sec:origins}. We argue below that the existence of the absorption troughs is \textit{not} the result of instrumental/algorithmic effects, as the feature is present in a variety of HETDEX spectral stacks and in other datasets. 

\subsection{Instrumental and Algorithmic Concerns}
\label{sec:instrumental}

\begin{figure}
    \centering

    \includegraphics[height=6cm]{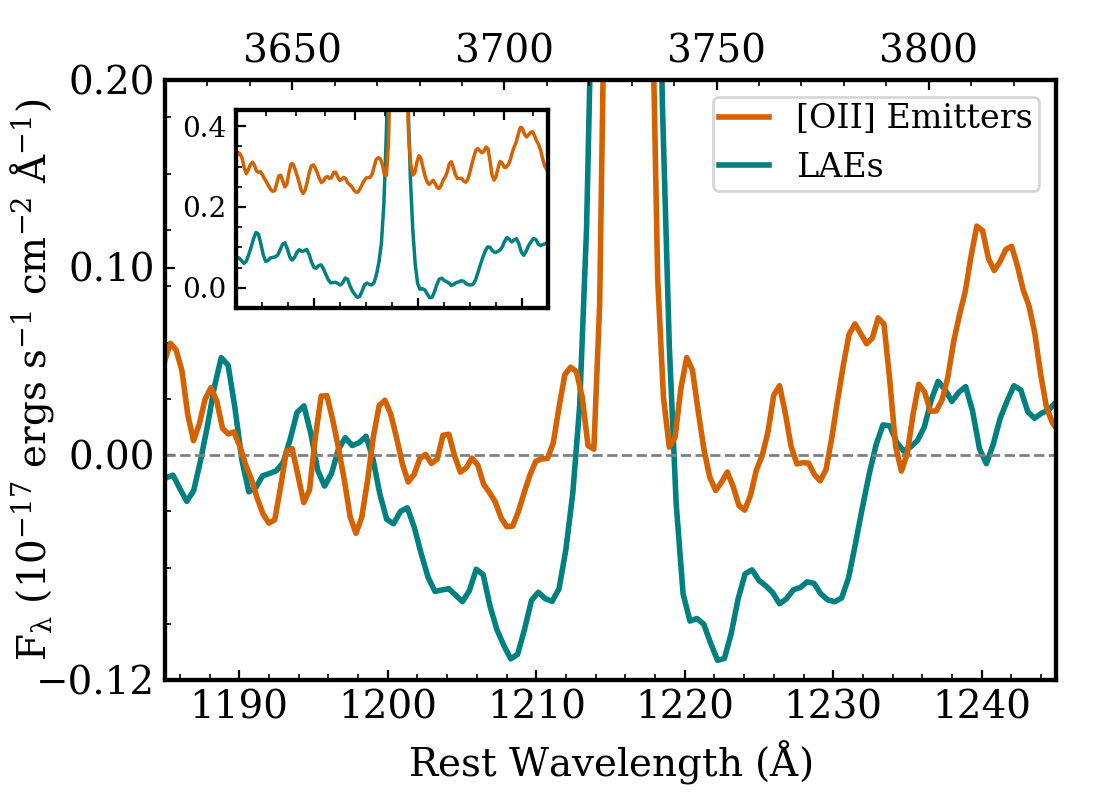}
    \caption{A stack of HETDEX LAEs alongside a stack of HETDEX \OII emitters. We compare emission lines of a similar brightness and/or equivalent width to ensure that the difference between them is intrinsic and not because of properties of an emission line detection. We chose the galaxies with $g$ or $r$ band counterpart magnitudes between 22 and 24, and line flux measurements $10-20 \times 10^{-17}~\mathrm{ergs~s^{-1}~cm^{-2}}$. Each stack contains 100-250 spectra. Note that there are inherently few \OII and LAE detections that are similar in line brightness due to cosmological dimming. The inset displays the \OII and LAE stacks, and the main panel overlays the profiles (continuum subtracted). Both the inset and main panel are plotted on the same wavelength scale. While the stack of LAEs has absorption troughs, the stack of \OII emitters does not.} 
    \label{fig:oii v lae}
\end{figure}

The absorption troughs in Figure \ref{fig:hpsc stack} are significantly broad, sharp and, concerningly, negative. Before we can begin to analyze the shape, evolution, and/or physical origins of the troughs, we must first conclude that they are not due to instrumental or algorithmic effects in HETDEX calibration and data processing. Sources of a systematic error we consider are 1) sky subtraction residuals, 2) problems with the detectors, 3) extended \lya emission, and 4) odd and plentiful false positive behavior.

The HETDEX sky subtraction algorithm assumes that the instrumental resolving power is the same over all fibers in an amplifier. A systematic trend in the resolving power of the fibers from the center to the edge of the detector could create correlated residuals. For example, subtraction of a bright emission line, if not properly modeled across the detector, could create false absorption wings. However, this residual would result in wings of about one wavelength pixel, while the trough widths we observe are roughly 35 pixels wide. Furthermore, if this residual were the cause, a brighter line would create deeper absorption wings, which is not the case (see Section \ref{sec:hetdex troughs}). More importantly, this effect would manifest itself in any emission line, and we do not observe the troughs in the detections of [\ion{O}{2}] $\lambda 3727$ or other emission lines detected in HETDEX (see Figure \ref{fig:oii v lae}).  

Another possibility lies in the detector electronics, where a charge transfer could cause a systematic oversubtraction. This effect is signal dependent, detector dependent, asymmetric in wavelength, and far too weak to affect approximately 35 pixels similarly on the 156 detectors. As stated above, this would be present in all detected emission lines, not just \lyaa. 

Alternatively, the \lya absorption troughs observed in the stacked spectra of LAEs may be the manifestation of faint, extended \lya emission distributed over most of an IFU's $50\arcsec$ field-of-view.  If such halos exist, the local sky estimator employed by HETDEX could cause the line profile of a LAE's halo to be subtracted from the profile of its core. This could produce the profile seen in Figure~1.  However, while extended \lya emission is present in HETDEX data \citep{Lujan_2022,Lujan_Bowman_2022}, troughs exist when using local sky subtraction \textit{and} when using full-field sky subtraction. The full-field sky is measured from all of HETDEX's 34,944 fibers spread over a 21\arcmin~field-of-view. This is far beyond the typical extent of \lya emission \citep{Wisotzki_2018}. Although there is a small amount of self-subtraction from \lya halos in the local sky, the effect is minimal, and the depth of the troughs changes only by $\sim$5\% when using local versus full-field sky subtraction. Assuming significantly extended \lya emission out to $\sim 1$ cMpc \citep{Martin_2023}, the emission would have to be significantly brighter/broader at large radii where the sky is measured than on the LAE. If we instead stack spectra that have \textit{not} been sky subtracted, the troughs around \lya are still present. 

Lastly, the \lya line profile of Figure~1 could be due to an overwhelming number of \lya false positives, each with the shape shown in Figure \ref{fig:hpsc stack}. False positives arise when random noise is interpreted as a low $S/N$ emission line. As presented in \cite{Cooper_2023}, the current false positive rate for detections with an $S/N$ of 
$>$ 5.5 is less than 15\%.  This rate is far too low to create the observed absorption troughs. Moreover, since there is no argument for the false positives consistently having the shape of spectral feature, we exclude the possibility that false positives are the cause of these troughs.

We conduct a number of additional tests to eliminate the possibility of a systematic issue within HETDEX's calibration. These tests include stacking spectra of empty fibers, limiting the LAE sample to objects in specific amplifiers, and at varying S/N threshold for \lya detections. As a basic test, we stack spectra from fibers chosen randomly spatially and in wavelength and, as expected, find no troughs. When stacking on LAEs, the troughs are consistent across amplifiers and are present with increasing line luminosity. The troughs, while not visible in an individual spectrum, appear in any random combination of $\sim100$ or more LAEs.  

The most convincing evidence that these absorption wings are not an artifact of data reduction and/or our stacking algorithm is that the troughs do \textit{not} appear in the stacks of HETDEX \OII emitters. \OII $\lambda 3727$ is detected in HETDEX roughly as frequently as \lyaa, and the spectrum of an [\ion{O}{2}] galaxy is calibrated in exactly the same way as that of an LAE\null. Figure \ref{fig:oii v lae} compares a stack of HETDEX \OII emitters to a stack of LAEs. For the sake of consistency, the spectra are chosen to have counterparts of a similar magnitude and similar line fluxes. Additionally, a possible continuum over-subtraction would be more evident in stacks of faint \OII emitters. The stack of LAEs shows the absorption troughs, while the stack of \OII emitters does not.

\subsection{\texorpdfstring{\lyaa}{Lyman Alpha} Troughs in Other Datasets}
\begin{figure*}
    \centering
    \includegraphics[height=5.9cm]{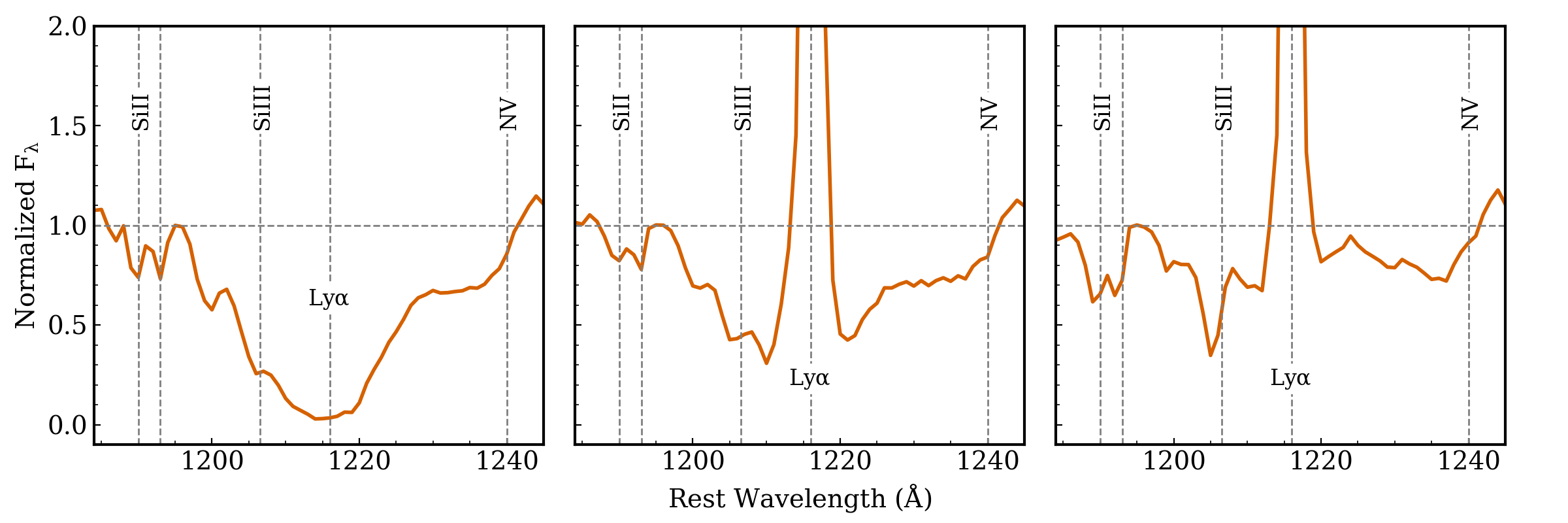}
    \caption{Spectral stacks of the galaxies in the CLASSY sample normalized to their continuua. \textit{Left panel}: A stack of the 19 CLASSY galaxies that exhibit pure \lya absorption. Various other absorption features can be seen, such as \ion{N}{5} and \ion{Si}{2}. \textit{Middle panel}: A stack of four CLASSY galaxies, each individually appearing to have \lya emission sitting in an absorption trough. The absorption redward of the \lya emission appears slightly wider than that observed in HETDEX. \textit{Right panel}: A stack of all 22 CLASSY galaxies that have \lya emission. There is absorption from the \ion{N}{5} P Cygni profile near \lyaa, but an obvious \lya absorption profile is not apparent.}
    \label{fig:classy stacks}
\end{figure*}

\begin{figure}
    \centering
    \includegraphics[height=6cm]{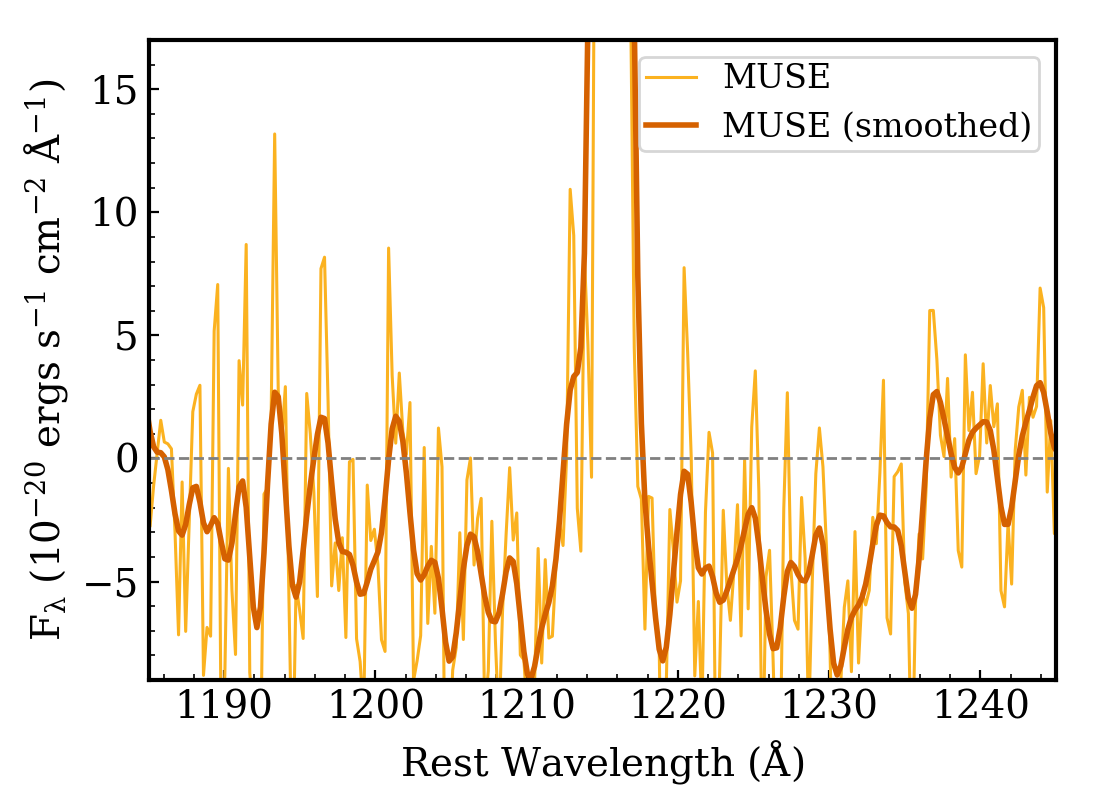}
    \caption{A stack of 169 MUSE-Wide LAE spectra selected at $2.9 < z < 4.3$ that have comparatively lower \lya line fluxes. The unsmoothed stacked spectrum is indicated in yellow, and the spectrum smoothed by a Gaussian kernel of three pixels is shown in orange. As with HETDEX stacks, we observe \lya absorption troughs on either side of the emission, and we also find that these troughs are negative. These troughs more closely resemble the ``boxy'' profiles we see in the HETDEX stack.}
    \label{fig:muse stack}
\end{figure}

Two public datasets applicable to HETDEX are CLASSY \citep{Berg_2022} and MUSE-Wide \citep{Bacon_2010, Urrutia_2019}. The CLASSY data explore 45 nearby ($z < 0.2$) galaxies that are chosen to be similar to high redshift LAEs. The MUSE-Wide sample overlaps HETDEX in redshift range and galaxy properties. 

The CLASSY LAEs have \lya luminosities $36.8 <$ Log$(L_{\mathrm{Ly}\alpha}) < 42.8$. After removing four galaxies from the CLASSY dataset that have their \lya profile contaminated by \lya or \ion{O}{1} from the Earth's atmosphere, we have 41 galaxies that we divide visually into three subsamples: 19 with \lya in absorption, four with \lya emission sitting in an absorption well like those in HETDEX, and 22 with \lya in emission. To compare to what we observe in HETDEX spectra, we rebin and convolve the CLASSY medium resolution spectra ($\sim 0.03$ \AA~pixel$^{-1}$) to HETDEX resolution using a 2~\AA~pixel$^{-1}$ kernel, de-redshift the wavelengths using the galaxy's systemic redshift according to \cite{Berg_2022}, fit a spline to their continua, and normalize to this fit.

Figure \ref{fig:classy stacks} depicts the three stacks of CLASSY spectra. The leftmost panel shows the 19 galaxies with \lya absorption, the middle panel depicts the 4 CLASSY spectra with \lya emission and absorption, and the rightmost panel represents the 22 galaxies with \lya emission. A number of additional absorption and emission features are present, such as \ion{Si}{2} $\lambda \lambda$1090, 1093 and \ion{N}{5} $\lambda \lambda$1238, 1242.  The shape of the middle profile differs from the HETDEX stack, as it more closely follows a typical Voigt profile, rather than the ``boxy'' troughs seen in Figure \ref{fig:hpsc stack}. For these particular CLASSY profiles, \cite{Hu_2023} reports that they are indicative of an inhomogeneous ISM with outflows and can be modeled by splitting the \ion{H}{1} covering factor between low and high density channels. Additionally, while several \lya absorption profiles in CLASSY are saturated, none are negative.  

Galaxies in the MUSE-Wide survey have redshifts at $2.9 < z < 6.3$. To compare to typical HETDEX LAEs, we first select the galaxies with $z \lesssim 4.3$, then we further select galaxies with \lya line fluxes in the fainter 55\% of the MUSE sample's flux distribution ($< \sim3\times 10^{-17}$ \escma). Both cutoffs are chosen roughly to reproduce the HETDEX \lya flux distribution while keeping $\sim 200$ individual spectra in each stack. We de-redshift the wavelengths of each spectra using the reported redshift from MUSE, but keep the flux densities in the observed frame. The stack of 169 MUSE-Wide galaxies is shown in Figure \ref{fig:muse stack}. As with HETDEX, there are absorption troughs around \lya emission, and, interestingly, the troughs are negative. We did not apply the corrections to convert the fluxes to the rest frame, nor did we continuum subtract.

The troughs seen in the CLASSY stack are consistent with absorption from the ISM. Those seen in MUSE-Wide are similar to what we observe in HETDEX, both in their negative flux values and ``boxy'' shape. These types of troughs are the focus of this paper. The physical origin of the troughs will be discussed further in Section \ref{sec:origins}. 

\subsection{\texorpdfstring{\lyaa}{Lyman Alpha} Troughs in HETDEX}
\begin{figure*}
    \centering
    \includegraphics[height=5.8cm]{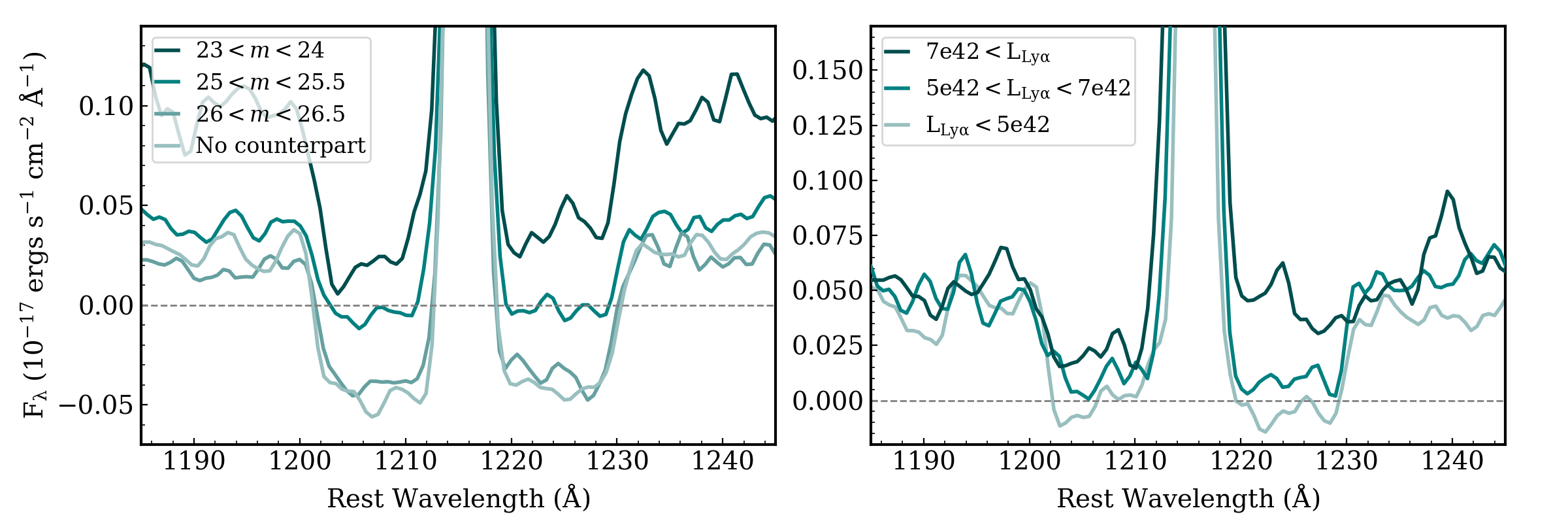}
    \caption{\textit{Left panel:} Stacks of HETDEX LAEs in counterpart magnitude bins. We restricted these subsamples to have $z \sim 2.5$ and $5 < S/N < 8$. The troughs themselves are present in each stack, although are not negative in stacks of brighter LAEs. This behavior is likely due to higher  continuum levels. \textit{Right panel:} Stacks of HETDEX LAEs in three \lya luminosity bins. These galaxies are selected to have counterpart magnitudes between 24 and 25 and $z \sim 2.5$. The strongest \lya absorption troughs appear in the stacks of galaxies with the faintest \lya emission. In the highest luminosity bin, the red trough is entirely filled in by \lya in emission. For both panels, there are roughly 200-1000 spectra contributing to the stacks.}
    \label{fig:hetdex mag lum stack}
\end{figure*}

\label{sec:hetdex troughs}
With hundreds of thousands of available LAE spectra, we stack according to host galaxy properties to investigate if the troughs change with the subsample. We first determine whether the absorption troughs are unique to LAEs that are significantly brighter or fainter in the continuum than the typical HETDEX LAE. About 25,000 HETDEX LAEs have confirmed counterparts via ELiXer cross-matching to various catalogs in the $r$ and/or $g$ band. We restrict this sample to the $2.4 < z < 2.6$ redshift range  to mitigate any redshift evolution and stack galaxies according to apparent magnitude in either band. The wavelengths of the spectra are shifted to the rest frame, while the fluxes remain in observed frame, as any multiplicative correction for cosmological dimming will deepen the artificially negative troughs. The stacks of these galaxies as well as those without a counterpart are displayed in the left panel of Figure \ref{fig:hetdex mag lum stack}. The troughs are present in LAEs with and without counterparts, and are not unique to galaxies of a certain magnitude. However, whether or not the troughs are negative does change, likely because our over-subtraction of sky flux does not push the troughs negative in LAEs with more underlying continuum flux. 

As a second experiment, we again confine our analysis to LAEs in the redshift range $2.4 < z < 2.6$, but we now fix the $g$ and $r$ continuum magnitudes ($24 < m < 25$),  while sub-dividing the sample by \lya line luminosity.   The troughs in the right panel of Figure \ref{fig:hetdex mag lum stack} are not negative due to the bright continuum level of this sample. The stack of galaxies with the faintest \lya emission has the strongest troughs, while the stacks of galaxies with more \lya emission have slightly weaker troughs. In the highest luminosity bin, the redward trough has been filled in by \lya emission. This effect resembles the asymmetry of the troughs seen in Figure \ref{fig:hpsc stack}, as slightly higher S/N spectra ($S/N > 5.5$) contribute to the stack. This trend with luminosity lends itself to the idea that these troughs do not arise due to instrumental/electronics issues, as more flux from the emission line would create deeper troughs. 

While we currently have the capabilities to stack as a function of redshift, we reserve this analysis for a future paper. Until we apply a physical model, the depths of the troughs will be artificially enhanced by the corrections to the rest frame. For the remainder of this paper, we focus on developing a physical interpretation of the negative troughs.

\section{Physical Model of \texorpdfstring{\lyaa}{Lyman Alpha} Troughs}
\label{sec:origins}

A physical model of the absorption troughs must explain three primary observables:  the depth of the absorption troughs, the width of the troughs, and the fact that they go negative with our sky subtraction. To begin investigating the physical origins of the \lya troughs, we note their similarity to DLAs. 

In DLAs, background light from a bright source such as a quasar is considerably absorbed at the \lya transition by hydrogen from intervening galaxies or protogalaxies both in the core of \lya and in the damped wings of the line \citep{Wolfe_1995}. The high \ion{H}{1} column density ($N$(\ion{H}{1}) $\geq 10^{20}~\mathrm{cm}^{-2}$) and frequency of these absorbers account for a significant fraction of the neutral hydrogen content in the universe between $0 < z < 5$ \citep{Storrie-Lombardi_2000}. 

DLA systems that also have \lya emission from the foreground galaxy (such as the one presented in Figure 1 of \cite{Moller_2004}) are similar to the profile shapes we see in stacks of HETDEX LAEs.  While it is clear that these troughs are not absorption of a bright individual background galaxy such as a quasar, we propose a solution using a similar geometry. In the model described below, the source of the light in the background of the LAEs is dependent on the density of the nearby galaxy population.

\subsection{Background Light}
\label{sec:uvb}

\begin{figure}
    \centering
   
    \includegraphics[height=5.75cm]{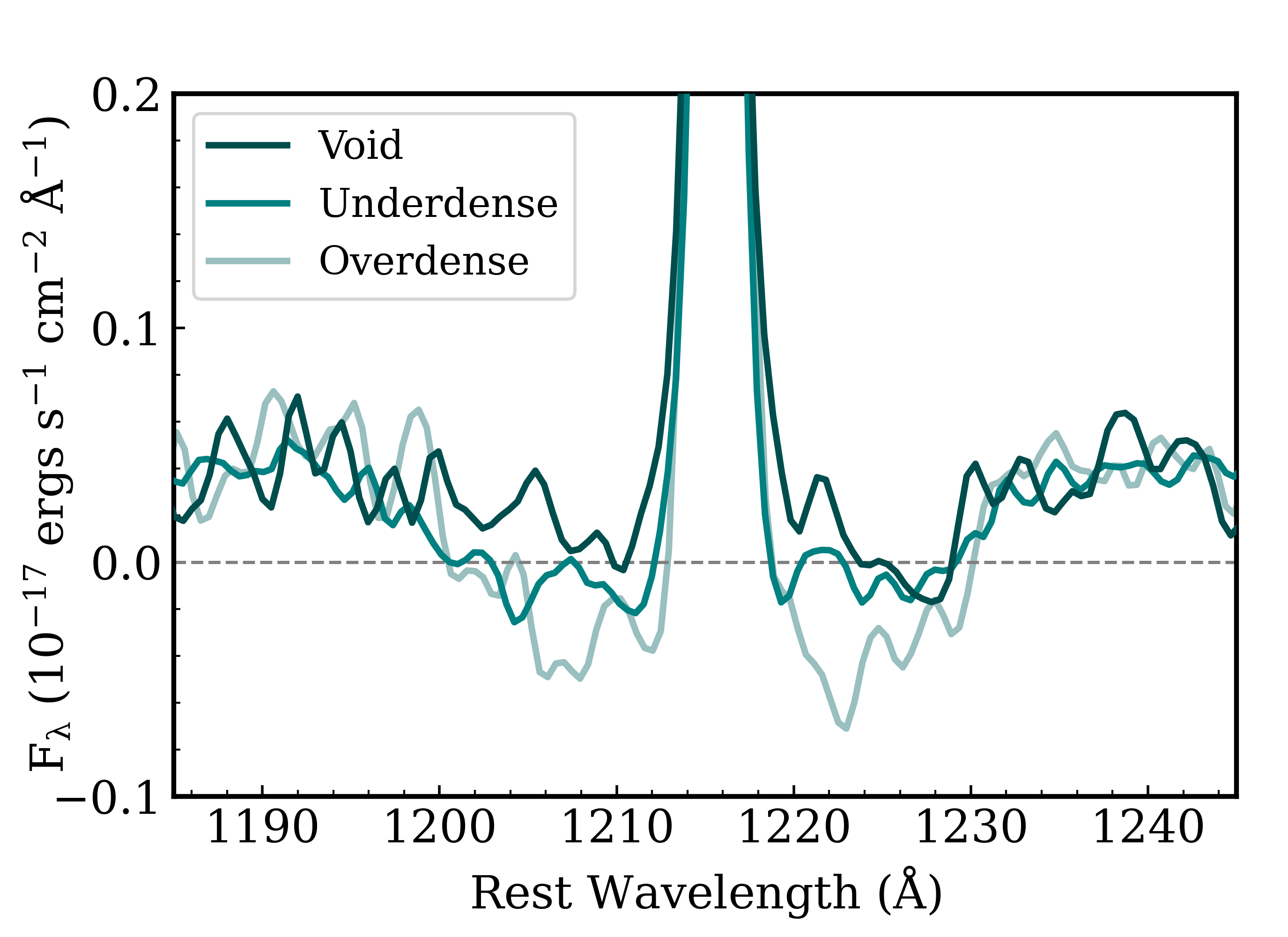}
    \caption{Stacks of galaxies that reside in over, under, and very under-dense fields using the luminosity function normalization as a proxy for density. The widths of the bins are chosen to have $\sim80-200$ contributing LAEs, with a similar number of spectra contributing to each stack. The troughs are weakest in very under-dense fields and strongest in over-dense fields. The troughs in the over-dense stack are the most negative, despite all three bins having similar continuum levels.}
    \label{fig:hetdex overden stack}
\end{figure}

The integrated light from galaxies behind the LAEs provides a spectrum of background radiation. Theoretically, there has been significant effort in modeling the radiative transfer of UV emission from AGN and star-forming galaxies through the IGM \citep[e.g.][]{Miralda_Escude_1990, Giroux_1996, Haardt_Madau_1996, Faucher_Giguere_2009, Haardt_Madau_2012}, which produces an evolving UV background (UVB). 

Observationally, redshifted photons from all sources throughout cosmic history have been observed across the electromagnetic spectrum as extragalactic background light (EBL) \citep{Lauer_2022}.  These studies generally measure a background light that is representative of an average over the full sky. In our case, we require a more local background spectrum, although the spectrum should be similar in shape, since it is due to stars in the background galaxies. The flux of the background, however, will vary with the density of nearby galaxies.

LAE at $z \sim 3$ trace the large-scale clustering of galaxies, with a linear bias likely in the range between $\sim 1.4$ and $\sim 2.0$ \citep{Gawiser_2007, Guaita_2010, Kusakabe_2018}.  Since the VIRUS fiber diameter and median HETDEX image quality are both moderately large ($1\farcs 5$ and $1\farcs 7$). respectively), the individual individual spectra contain contributions well into the halos of $z \sim 2-3$ LAEs. Therefore, every HETDEX observation of an LAE must also contain diffuse flux from background sources both nearby and distant, as well as absorption at \lya by gas in the halo. Averaging over tens of thousands of LAEs will account for galaxies of all magnitudes, including bright neighbors. The absorption troughs only become significant when stacking, since the intensity of the background spectrum in the observed frame is significantly fainter than the typical noise in a single-fiber spectrum. In this picture, the troughs deepen with increasing density of background galaxies. Thus, the absorption around \lya should be strongest in over-dense fields. We can test this hypothesis by examining the behavior of the \lya emission versus environment. To do this, we start by determining the local galaxy luminosity function for every HETDEX pointing.

Each individual 21\arcmin\ field for HETDEX contains anywhere from $\sim$100 to $\sim$400 LAEs. These LAEs have redshifts from $1.88 < z < 3.52$; we can generate a luminosity function from each field with this pencil-beam geometry. We can then normalize these functions to the luminosity function formed from all $\sim$800,000 LAEs contained in the 3993 fields of HDR3 (Gebhardt et al., in prep, Jeong et al., in prep) to obtain a rough estimate of the relative galaxy density along each line-of-sight. Although we would prefer to use an estimate of environment that uses the galaxy density surrounding each LAE, rather than one based on a line-of-sight summation, the integrated approach is the best that can be done with the present data.

Using these luminosity function normalizations ($LF_{\mathrm{norm}}$), we define three environments: over-dense ($1.6 < LF_{\mathrm{norm}} < 1.61$), under-dense ($0.3 < LF_{\mathrm{norm}} < 0.4$), and void ($0 < LF_{\mathrm{norm}} < 0.28$) regions. The limits and bin widths are chosen only to ensure sufficient spectra to stack, as there are naturally more galaxies in over-dense fields. The stacked spectra of the galaxies within each bin are are presented in Figure \ref{fig:hetdex overden stack}. The troughs are barely visible in very under-dense regions and become more significant and negative with increasing field density. While more robust density calculations are needed for further analyses, the observed trend is consistent with our proposed background light absorption scenario.

\subsection{Over-Subtraction of Sky}
\label{sec:oversub}

\begin{figure}
    \centering

    \includegraphics[height=5.75cm]{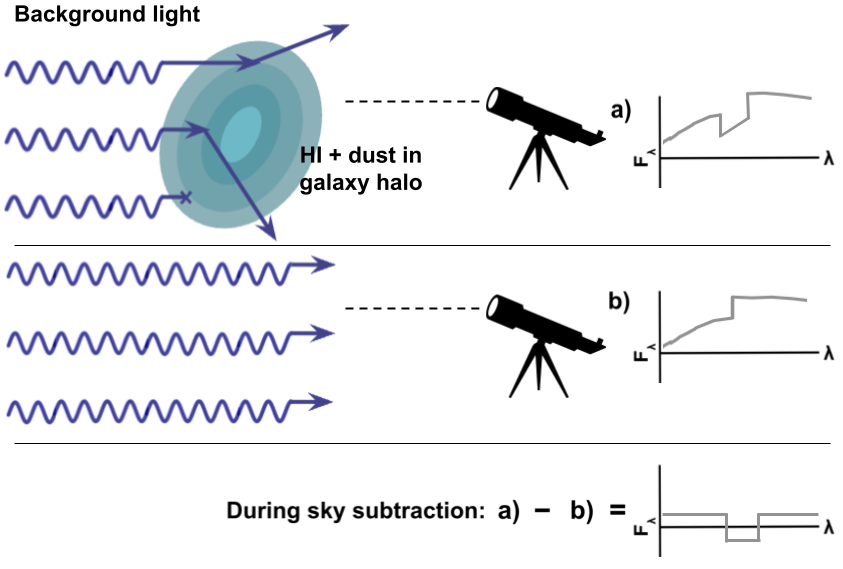}
    \caption{A simplified, 1D graphic depicting our proposed interpretation of the negative \lya troughs. In a), background UV light is absorbed at the wavelength of \lya by \ion{H}{1} gas in the halo of the LAE, creating an absorption well in the combined spectrum. The \lya emission from the foreground LAE is not shown for clarity. In b), the spectrum observed off source (empty sky) includes this UV background \textit{without} the \lya absorption. The shape of the background spectrum is as modeled by \cite{Haardt_Madau_2012}, while the shape of the absorption and LAE continuum are simple representations. During HETDEX's sky subtraction, b) is subtracted from a) creating the negative absorption seen in the bottom panel.}
    \label{fig:graphic}
\end{figure}

Negative troughs are consistent with the physical interpretation that typical LAEs absorb background light produced by nearby galaxies. The negative troughs arise because the flux at the wavelength of \lya is removed from a spectrum twice: first physically by an LAE and then again during the sky subtraction procedure. 

When the background photons at the resonance wavelength of \lya are scattered by \ion{H}{1} and absorbed by dust in a galaxy's halo, the contribution of the background at \lya is removed, creating the troughs like those in our stacked fiber spectra. When applying the sky subtraction (see Section \ref{sec:corrections}), we assume that the background sky is the same in the nearby fibers as it is in the fiber centered on the LAE. Since this assumption is \textit{not} correct due to the \ion{H}{1} absorption, we over-subtract at \lyaa. As a result, the troughs in the subtracted spectrum are pushed negative depending on the object's continuum level. The stack of MUSE spectra in Figure \ref{fig:muse stack} shows similar negative troughs; whether or not they are due to the same physical interpretation depends on the MUSE sky subtraction procedure. 

Figure \ref{fig:graphic} depicts a graphic of the over-subtraction process. In this figure, the top panel presents the full, on-target spectrum, the middle panel shows the spectrum off-target (i.e., our measurement of sky), and the bottom panel displays the sky-subtracted spectrum. Since we assume that the off-source sky is the same as the on-source sky, we over-subtract at \lya due to the absorption by \ion{H}{1} in the halo of the LAE. 

Because the background flux is faint, the over-subtraction at the wavelength of \lya on a per-fiber basis for an individual LAE is negligible. By stacking the spectra of hundreds to thousands of LAEs, the negative troughs become significant. In essence, a slight over-subtraction in the HETDEX data is indicative of local density enhancements and gas present in LAE halos. 
 
\subsection{\texorpdfstring{\lyaa}{Lyman Alpha} Profile Modeling}
\label{sec:model}

\begin{figure}
    \centering
  
    \includegraphics[height=6cm]{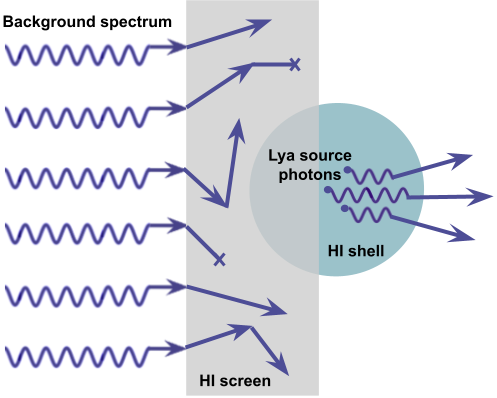}
    \caption{A graphic of the geometry used in modelling our proposed absorption scenario (note that this is an oversimplification of the physical model we suggest and serves only to illustrate the basic geometry). We place a continuum source that represents the average stacked ``background spectrum'' behind a screen of \ion{H}{1}. We then model a foreground ``LAE'' as a source of \lya photons surrounded by a less dense shell of \ion{H}{1} and partially embed it in the screen. The background photons are scattered/absorbed by the \ion{H}{1} screen, producing the absorption well. The foreground \lya source photons scatter in the less dense shell and escape along the line of sight for observation. The degree to which the ``LAE'' is embedded in the screen affects only the shape of the \lya emission component, which we set to reproduce the emission observed in our stacks. The overall profile is then the combination of the background absorption well and the foreground \lya emission.}
    \label{fig:graphic2}
\end{figure}

To further investigate our interpretation of the \lya absorption troughs, we create a phenomenological model of the \lya profile. We use RASCAS \citep[RAdiative SCattering in Astrophysical Simulations,][]{Michel_2020} to model the proposed background absorption scenario, with the goal of reproducing the observed absorption troughs in HETDEX. 

Using the HPSC-1 stack shown in Figure \ref{fig:hpsc stack}, we attempt to ``undo'' the over-subtraction caused by HETDEX sky subtraction. To each sky-subtracted, observed-frame LAE spectrum, we add a spectrum with the shape modeled by \cite{Haardt_Madau_2012} for the given redshift shifted in wavelength space to the observed frame. Although the spectrum from \cite{Haardt_Madau_2012} is integrated from the redshift of the LAE to infinity and does not account for local density variations, the source of the radiation is similar.  

However, for each individual LAE in the stacked spectrum, we do not know the background flux due to the density variations. Since our goal is to use this model to explain the over-subtraction, we apply the \cite{Haardt_Madau_2012} spectrum scaled such that the depths of absorption troughs in the stacked spectrum are no longer negative. This scaling is likely not correct due to the unaccounted for contribution of the underlying continuum of the LAEs. 

\begin{figure}
    \centering
  
    \includegraphics[height=6cm]{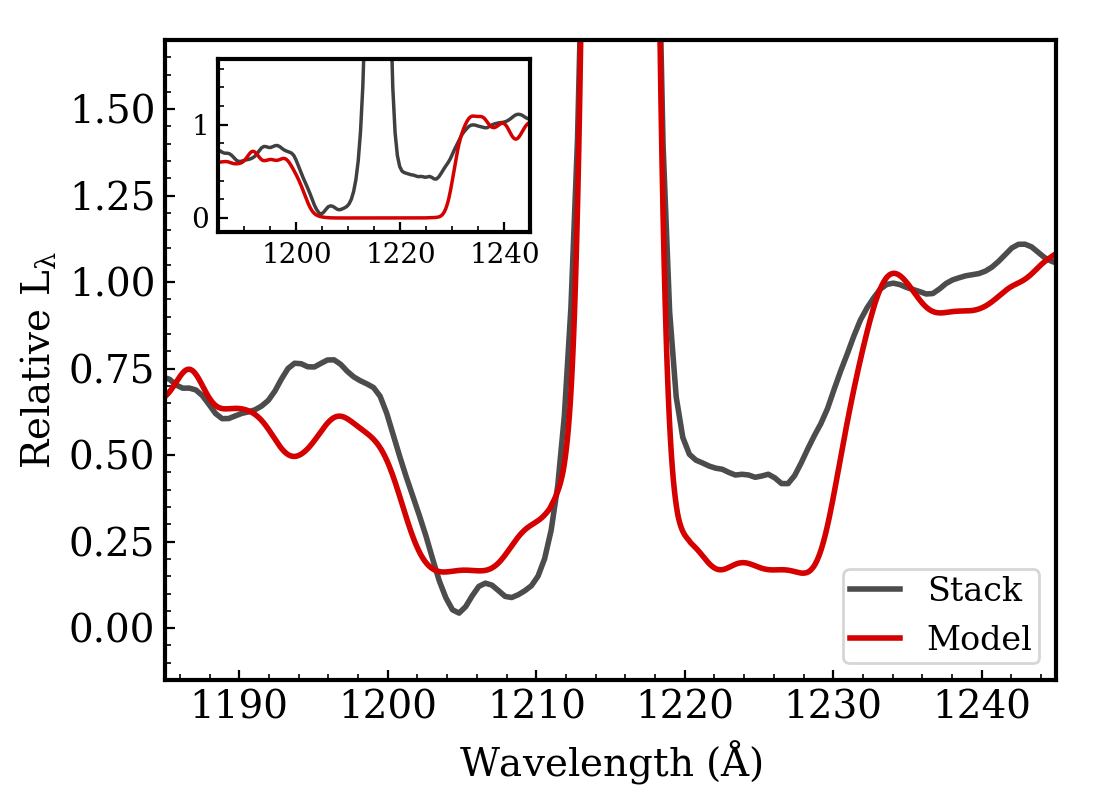}
    \caption{The \lya profile produced by our simulation of a background flux absorption scenario. The modelled profile is indicated in red and the HPSC-1 combined spectrum we fit to is plotted in gray. The upper left inset displays the profile fit to the absorption only. While the model does not simulate the continuum of the LAE and adopts a simple screen geometry for \ion{H}{1} gas and dust, it is able to reproduce a profile resembling the absorption troughs around \lyaa.} 
    \label{fig:model}
\end{figure}

This procedure creates the average combined background and LAE spectrum for HPSC-1 (see the black line in Figure \ref{fig:model}, plotted in normalized units). The troughs are no longer negative and the asymmetry of the profile has been enhanced due to the addition of the scaled, asymmetric \citep{Haardt_Madau_2012} background spectrum. This procedure also generates an average background spectrum, which is used as input to the simulation described below. We aim only to reproduce the troughs with physical geometry. Given our assumptions in this model, we do not attempt to quantify the intensity of the background or properties of the gas at this time.  

RASCAS is used to build a simple idealized model of our proposed scenario, with parameters chosen to reproduce the troughs seen in the HPSC-1 stack. We simulate a source of \lya photons at systemic velocity surrounded by a shell of \ion{H}{1}, with column density $N$(\ion{H}{1})$=10^{18} ~\mathrm{cm}^{-2}$ and a small density of dust ($n_{\mathrm{dust}} = 0.01~\mathrm{cm}^{-3}$) per cell at the systemic velocity of the LAE. Note, because we stack using \lya based redshifts, the \lya emission is centered at 0 \kmss. This \lya source is partially $\mathbf{(\sim 30\%)}$ embedded in a screen of \ion{H}{1} ($N$(\ion{H}{1})$ = 10^{19.5}~\mathrm{cm}^{-2}$) with $n_{\mathrm{dust}} = 0.005~\mathrm{cm}^{-3}$. To determine parameters for the \ion{H}{1} screen, we fit a single Voigt profile to the stack's absorption well (see the inset plot of Figure \ref{fig:model}). This fit yields a velocity offset of $-100$ \kms (blueward of the \lya emission) and a Doppler broadening parameter, $b$, of 1050 \kms which we adopt for the screen. We then model a continuum source with the spectrum and intensity of the average stacked background described above, also at systemic velocity. This continuum is placed behind the \ion{H}{1} screen. A graphic of this geometry is depicted in Figure \ref{fig:graphic2}. 

With this configuration, we create 35,000 photon packets emitted at 0 \kms to sample a Gaussian \lya profile and 15,000 to sample a background spectrum. We propagate all photons through the geometry described above with RASCAS. Each time the photon enters a cell, the optical depth of the \ion{H}{1} and dust in the cell is computed, and then the probability that this photon is scattered by \ion{H}{1} or absorbed by dust is calculated. If the photon is scattered, it is assigned a random direction for re-emission with a new calculated frequency. This procedure is repeated through the grid for each photon. The final profile includes all the photons that escape in a randomly chosen viewing direction. 

In this model, the background photons are scattered and absorbed by the dense screen, producing the broad absorption well. Because the \lya source is only partially embedded in the screen, the \lya photons scatter in the less dense shell and a large fraction manage to escape along the line of sight, producing the observed \lya emission. The geometric picture that the LAE is asymmetrically embedded in the \ion{H}{1} screen makes sense, as LAEs obscured by a sufficient amount of \ion{H}{1} (even with a low density of dust) would not be detected in emission. Thus, we are likely observationally biased towards LAEs with less material in the immediate foreground. In this case, there are also likely fewer galaxies in the foreground of the LAE than in the background, creating the local, anisotropoic background consistent with this model. The resulting profile is the sum of the foreground LAE and absorption of the local background. 

The red line in Figure \ref{fig:model} represents the spectrum generated by the parameters described above and is scaled in relative units. The inset panel depicts the profile of only the background and \ion{H}{1} screen component. This creates an underlying saturated absorption well, without the contribution of the \lya emission. Adding the embedded \lya photon source produces the combined emission and absorption profile. This model can produce symmetric, broad, and boxy troughs similar to those observed in HETDEX stacks. The red trough of the stack is shallower than in the model, i.e., similar to the luminosity trend in Figure \ref{fig:hetdex mag lum stack}. It is possible that this asymmetry in the troughs is influenced by high \lya luminosity spectra contributing to the HPSC-1 stack, although more complex models that include stellar continua are needed to investigate this further. Changing the density of dust and the degree to which the \lya emitting region is embedded in the \ion{H}{1} screen changes the shape of the \lya emission component, which alters the overall combined profile. We reserve the fine-tuning of these parameters for a future paper.  
 
If we assume the \lya emission of the HPSC-1 stack is redshifted from systemic by $\sim 250$ \kms \citep{Davis_2023_hpsc}, the absorbing gas of the fit model ($-100$ \kmss) is redshifted from the systemic velocity by $\sim 150$ \kmss. While there are several combinations of $b$ and $N$(\ion{H}{1}) that produce similar profiles, in order to reproduce the broad and boxy troughs seen in HETDEX, $b$ must be large. If we instead restrict $b < 200$ \kms to reflect typical conditions of the ISM, the modeled profile more closely resembles the stack of the four CLASSY galaxies with \lya absorption and emission (see the middle panel of Figure \ref{fig:classy stacks}). The \lya profiles in these CLASSY galaxies have been previously fitted in this manner \citep{Hu_2023}. 

The need for a much larger Doppler parameter to replicate the absorption troughs in the HETDEX stacks may indicate that the absorbing gas is not within the ISM (though some amount of LAE continuum absorption by the ISM likely contributes to the overall profile). The HPSC-1 stack is made using LAEs from different redshifts, thus any evolution of the troughs will be artificially smeared and affect the profile. 

\begin{figure}
    \centering
   
    \includegraphics[height=8.5cm]{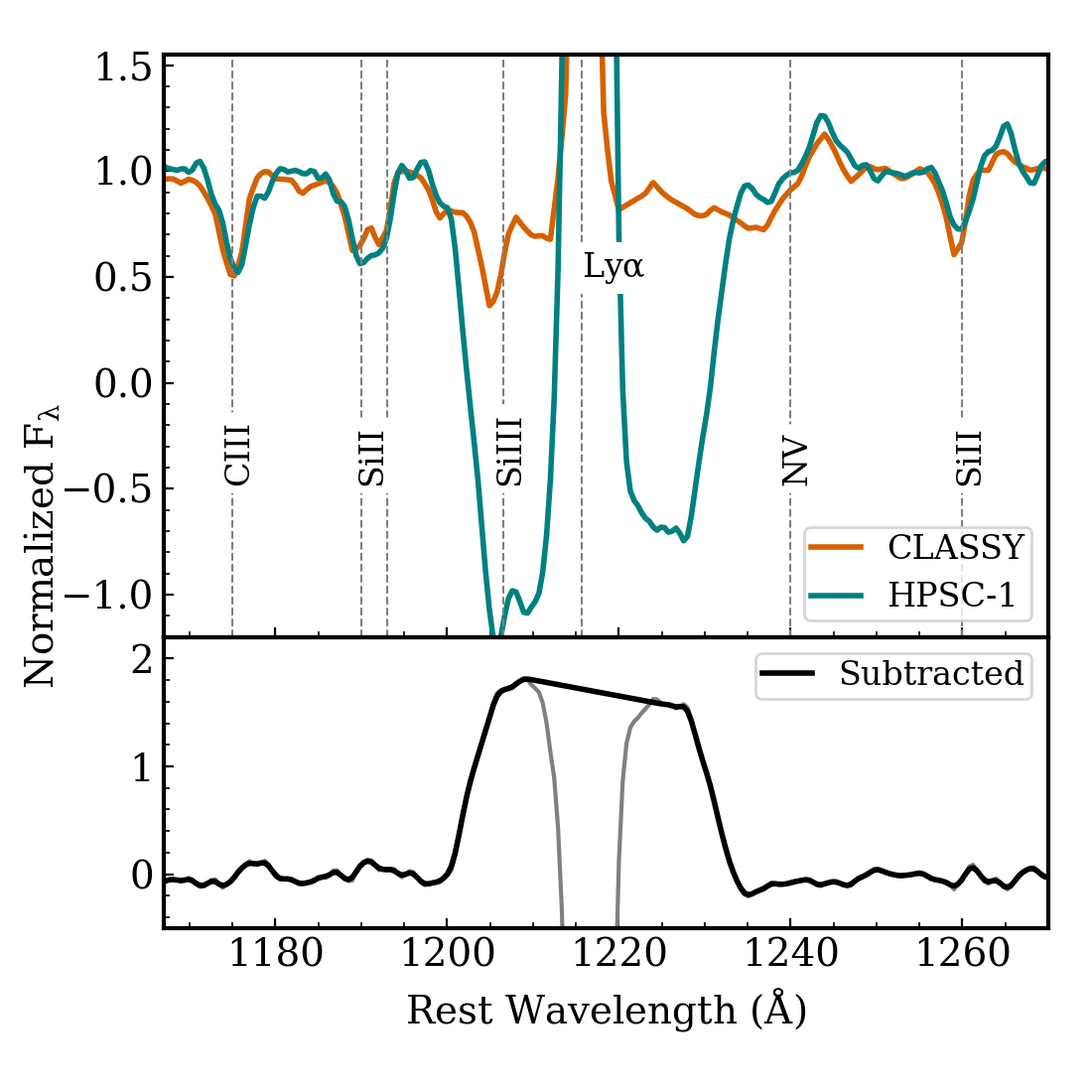}
    \caption{\textit{Upper panel:} The HPSC-1 stack compared to the stack of CLASSY LAEs convolved to the spectral resolution of VIRUS (see the rightmost panel of Figure \ref{fig:classy stacks}).  Both stacks have been normalized to their continuum. The HPSC-1 stack has also been shifted by 250 \kms to account for the \lya velocity offset from systemic. \textit{Lower panel:} The normalized HPSC-1 stack subtracted from the normalized CLASSY stack, with the \lya emission masked out. In the background absorption scenario, this profile reflects the absorption of the integrated flux from background sources by \ion{H}{1} in the LAE halo.}
    \label{fig:classy v hetdex}
\end{figure}

We can empirically determine the absorption profile by subtracting the HPSC-1 stack from our expectation for \lya using the CLASSY stack (see the right panel of Figure \ref{fig:classy stacks}). The continuum normalized HPSC-1 stacks are displayed in the top panel of Figure \ref{fig:classy v hetdex}, and the subtraction of the two is given in the bottom panel, masking the central \lya emission. The HPSC-1 stack has been shifted by 250 \kms to account for the \lya velocity offset from systemic. Effectively, the shape of this subtracted profile is representative of the typical absorption of background flux by \ion{H}{1} in the halo of LAEs. 

\section{Summary}
\label{sec:discuss}

The unprecedented number of LAE spectra in HETDEX provides an opportunity to explore the production and radiative transfer of \lya in the $1.9 < z < 3.5$ epoch of the universe. While an individual LAE spectrum lacks sufficient $S/N$ for in-depth measurements of the \lya profile or continuum level, a stack of LAEs can provide far more detail and reach flux limits impossible to otherwise attain. 

In this paper, we explored spectral stacks of up to 50k LAEs. Such data not only enable investigations of weak absorption features in the continua of these objects, but also the detailed analysis of the \lya line profiles. One of the main obvious features consistent across these stacks is the deep absorption around the wavelength of \lyaa. While these troughs reach flux density levels below zero, we argue that their existence is not an artifact of the data, but are a real feature of LAE spectra.

We propose a model wherein flux from nearby background sources is absorbed by the \ion{H}{1} gas around the LAE, in a geometry similar to that of DLAs. This flux is over-subtracted around the wavelength of \lya during our sky-subtraction procedure, creating negative troughs, whose depths depend on the local density enhancements. We use field number counts as a proxy for the density around the LAE, and stack galaxies in over- and under-dense fields. We find that the strengths of the troughs increase with density. A simple radiative transfer simulation created to model this geometry is able to produce troughs similar to those seen in HETDEX stacks. 

HETDEX will eventually have more than one million LAE spectra from $1.88 < z < 3.52$, and we will be able to combine spectra in various ways. Of particular interest in future work is the evolution of the absorption troughs with redshift and environment. To explore the spatial distribution of background absorption, we plan to stack spectra in annuli around the LAEs as well as between LAE pairs. With more in-depth simulations, we aim to determine the intensity of the LAEs' local background and quantify the properties and evolution of the absorbing gas.

\acknowledgments

HETDEX is led by the University of Texas at Austin McDonald Observatory and Department of Astronomy with participation from the Ludwig-Maximilians-Universit\"at M\"unchen, Max-Planck-Institut f\"ur Extraterrestrische Physik (MPE), Leibniz-Institut f\"ur Astrophysik Potsdam (AIP), Texas A\&M University, The Pennsylvania State University, Institut f\"ur Astrophysik G\"ottingen, The University of Oxford, Max-Planck-Institut f\"ur Astrophysik (MPA), The University of Tokyo, and Missouri University of Science and Technology. In addition to Institutional support, HETDEX is funded by the National Science Foundation (grant AST-0926815), the State of Texas, the US Air Force (AFRL FA9451-04-2-0355), and generous support from private individuals and foundations.

Observations were obtained with the Hobby-Eberly Telescope (HET), which is a joint project of the University of Texas at Austin, the Pennsylvania State University, Ludwig-Maximilians-Universit\"at M\"unchen, and Georg-August-Universit\"at G\"ottingen. The HET is named in honor of its principal benefactors, William P. Hobby and Robert E. Eberly.

VIRUS is a joint project of the University of Texas at Austin, Leibniz-Institut f\"ur Astrophysik Potsdam (AIP), Texas A\&M University (TAMU), Max-Planck-Institut f\"ur Extraterrestrische Physik (MPE), Ludwig-Maximilians-Universit\"at Muenchen, Pennsylvania State University, Institut fur Astrophysik G\"ottingen, University of Oxford, and the Max-Planck-Institut f\"ur Astrophysik (MPA). In addition to Institutional support, VIRUS was partially funded by the National Science Foundation, the State of Texas, and generous support from private individuals and foundations.

The authors acknowledge the Texas Advanced Computing Center (TACC) at The University of Texas at Austin for providing high performance computing, visualization, and storage resources that have contributed to the research results reported within this paper. URL:http://www.tacc.utexas.edu

The Institute for Gravitation and the Cosmos is supported by the Eberly College of Science and the Office of the Senior Vice President for Research at the Pennsylvania State University.

KG acknowledges support from NSF-2008793.  EG acknowledges support from NSF grant AST-2206222. ASL acknowledges support from Swiss National Science Foundation. 
HK and SS acknowledges the support for this work from NSF-2219212. EK and SS are supported in part
by World Premier International Research Center Initiative (WPI Initiative), MEXT, Japan.

\facility{HET}

\software{Astropy \citep{astropy:2018}, NumPy \citep{numpy}, SciPy \citep{SciPy}, Matplotlib \citep{matplotlib}, EliXer \citep{Davis_2023}}


\clearpage

\bibliography{lya.bib}


\end{document}